\documentclass[aps,floatfix,amsmath,nofootinbib,amssymb,superscriptaddress]{revtex4}
\usepackage{overpic}
\usepackage{amssymb}
\usepackage{indentfirst}
\usepackage{feynmf}
\usepackage{slashed}
\usepackage{cases}
\usepackage{color}
\usepackage{multirow}
\usepackage{epstopdf}
\usepackage{graphicx,color,bm}
\usepackage{epstopdf}

\usepackage[colorlinks,
            citecolor=blue,
            anchorcolor=red,
            menucolor=red,
            linkcolor=red,
            filecolor=red,
            runcolor=red,
            urlcolor=blue,
            frenchlinks=red]{hyperref}

\begin{document}

\title {Transverse momentum weighted Sivers asymmetries in SIDIS and Drell-Yan processes at COMPASS}

\author{Shi-Chen Xue}
\affiliation{School of Physics and Microelectronics, Zhengzhou University, Zhengzhou, Henan 450001, China}
\author{Shuailiang Yang}
\affiliation{School of Physics and Microelectronics, Zhengzhou University, Zhengzhou, Henan 450001, China}
\author{Xiaoyu Wang}
\email{xiaoyuwang@zzu.edu.cn}
\affiliation{School of Physics and Microelectronics, Zhengzhou University, Zhengzhou, Henan 450001, China}
\author{De-Min Li}
\email{lidm@zzu.edu.cn}
\affiliation{School of Physics and Microelectronics, Zhengzhou University, Zhengzhou, Henan 450001, China}
\author{Zhun Lu}
\email{zhunlu@seu.edu.cn}
\affiliation{School of Physics, Southeast University, Nanjing, Jiangsu 211189, China}

\begin{abstract}
We investigate the transverse momentum weighted Sivers asymmetries in the processes with transversely polarized proton target, including the $\frac{P_{hT}}{zM_p}\sin (\phi_h-\phi_S)$ weighted asymmetry in charged hadron production in semi-inclusive deeply inelastic scattering~(SIDIS) and the $\frac{q_T}{M_p}\sin (\phi_S)$ weighted asymmetry in $\pi^- p$ Drell-Yan process.
Due to the integration over the transverse momentum, the weighted asymmetries can be expressed as the product of the transverse-moments of the transverse momentum dependent (TMD) parton distribution functions (PDFs) and fragmentation functions (FFs).
Using the parametrization for the Sivers function of the proton, the unpolarized PDFs of proton and pion, and the unpolarized FF of charged hadron, we present the numerical calculation for weighted Sivers asymmetries in SIDIS and Drell-Yan processes at the kinematics of COMPASS, and compare them with experimental data.
We find that our prediction on the weighted Sivers asymmetry in SIDIS process is in agreement with
the recent COMPASS measurement.
Due to the relatively large uncertainties of the preliminary data from COMPASS Drell-Yan program, high precision experimental data are needed to test the sign change property of the Sivers function between SIDIS and Drell-Yan processes and to constrain the sea quark Sivers function.

\end{abstract}

\maketitle

\section{INTRODUCTION}

Since the European Muon Collaboration performed the measurement of the fraction carried by the internal quark of the proton spin~\cite{EuropeanMuon:1987isl,EuropeanMuon:1989yki}, the significant deviation from the theoretical prediction based on the quark model has inspired a large number of experimental and theoretical research for the proton spin structure. Understanding the spin structure of the nucleon has become one of the main goals in QCD and hadronic physics.
Normally, the 3-dimensional partonic structure of the nucleon is described by TMD PDFs.
Sivers function $f_{1T}^{\perp}$~\cite{Sivers:1989cc} is one of the eight TMD PDFs at leading twist, which is a time-reversal odd  (T-odd) distribution function that denotes the asymmetric distribution of unpolarized quarks inside a transversely polarized nucleon.
Arising from the correlation between the internal quark transverse momentum and the nucleon transverse spin, the Sivers function manifests novel spin structure of hadrons within the twist-2 approximation of QCD parton model. Because of its T-odd property, the Sivers function and its chiral-odd partner the Boer-Mulders function are forbidden by the naive time-reversal invariance of QCD, thus the very existence of the T-odd distribution functions was not so obvious.
However, the situation was changed after the calculations in Refs.~\cite{Brodsky:2002cx,Brodsky:2002rv,Boer:2002ju},
which show that the T-odd distributions can actually survive using spectator model calculations incorporating gluon exchange between the struck quark and the spectator.
The time-reversal-invariance argument was reexamined in Ref.~\cite{Collins:2002kn}, which shows that the gauge-link in the operator definition of the correlator guarantees the T-odd distribution functions to be nonzero.
Particularly, the presence of the gauge-link indicates that the Sivers function and the Boer-Mulders function have opposite sign between semi-inclusive deeply inelastic scattering~(SIDIS) and Drell-Yan processes~\cite{Brodsky:2002rv,Brodsky:2002cx,Collins:2002kn}, a significant prediction by QCD. The verification of this sign change is one of the most fundamental tests of our understanding of the QCD dynamics and the factorization scheme, and it is also the main pursue of the existing and future Drell-Yan facilities.

The transverse single-spin asymmetries (TSSAs) related to the Sivers function in high energy scattering processes turn out to be important experimental tools to investigate the information of Sivers function. During the past two decades, there are plenty of experimental measurements in SIDIS and Drell-Yan processes. Measurements in SIDIS process have been made by the HERMES Collaboration~\cite{HERMES:2004mhh,HERMES:2009lmz,HERMES:2020ifk}, COMPASS
Collaboration~\cite{COMPASS:2008isr,COMPASS:2010hbb,COMPASS:2012dmt,COMPASS:2016led}, and JLab Collaboration~\cite{JeffersonLabHallA:2011ayy, JeffersonLabHallA:2014yxb}. Meanwhile, the first measurement of TSSAs from Sivers function in the pion-induced Drell-Yan process is reported by COMPASS~\cite{COMPASS:2017jbv}. Measurement of TSSAs of weak boson produced in transversely polarized proton-proton collisions has also been performed by STAR experiment at RHIC~\cite{STAR:2015vmv}. The TSSAs data were adopted by several
groups~\cite{Anselmino:2005ea,Efremov:2004tp,Collins:2005ie,Vogelsang:2005cs,Anselmino:2008sga,Anselmino:2012aa,Bacchetta:2011gx,Echevarria:2014xaa,Anselmino:2016uie,Martin:2017yms,Boglione:2018dqd,Bury:2021sue} to extract the Sivers function from parametrization. From the theoretical side, the quark Sivers function has been intensively studied by various models, such as the spectator model~\cite{Brodsky:2002cx, Boer:2002ju, Bacchetta:2003rz}, the light-cone quark model~\cite{Lu:2004hu, Pasquini:2010af}, the non-relativistic constituent quark model~\cite{Courtoy:2008vi}, and the MIT bag model~\cite{Yuan:2003wk,Courtoy:2008dn}. Very recently, the sea quark Sivers function has been estimated from the light-cone wave function in Ref.~\cite{Luan:2022fjc}. Although there are extensive studies about Sivers function by theoretical studies, experimental measurements, and phenomenological analyses, what we know about Sivers function is still limited. One of the reasons is that the transverse momentum dependence of the Sivers function is tricky theoretically. The transverse spectrum in the large transverse momentum~(the transverse momentum of the final-state hadron in SIDIS process and the transverse momentum of the final-state dilepton in Drell-Yan process) region of the processes is expected to be described by the fixed order perturbative calculation of QCD, while in the small transverse momentum region, the transverse spectrum is described by the resummation of the soft-gluon. Thus, it is relatively difficult to describe the physical observables because of the complicated transverse momentum dependent effects.

Choosing proper transverse momentum dependent weighting function $W$ with the transverse momentum integrated out, the authors in Refs.~\cite{Kotzinian:1995cz,Boer:1997nt} proposed the idea of transverse momentum weighted asymmetry, which has been widely applied to SIDIS process~\cite{Pasquini:2011tk, Anselmino:2011ay,xue:2021svd,Bacchetta:2010si}
and Drell-Yan process~\cite{Sissakian:2005yp,Lu:2011qp,Wang:2017onm,Bacchetta:2010si,Liu:2021boj}.
Weighted asymmetry can be expressed as the product of collinear functions instead of the complicated convolution of the TMD functions considering the TMD effects. 
COMPASS experiment has the unique advantage to explore the sign change of the Sivers function since it has almost the same setup for SIDIS and Drell-Yan process, which may reduce the uncertainties in the extraction of the Sivers function from the two kinds of measurements.
Recently, COMPASS Collaboration has reported the measurements of weighted Sivers asymmetry in charged hadron produced SIDIS process in Ref.~\cite{COMPASS:2018ofp}. The preliminary measurement at COMPASS Drell-Yan program was also presented in Ref.~\cite{Longo:2019bih}, in which $\pi^-$ beam was scattered off the transversely polarized NH$_3$ target. These data provide an ideal opportunity to verify the sign change of T-odd PDFs through simultaneous analysis of the weighted Sivers asymmetries in both SIDIS and Drell-Yan processes. 

In this work, we first numerically calculate the weighted $\frac{P_{hT}}{zM_p}\sin (\phi_h-\phi_S)$ asymmetry in charged hadron produced SIDIS process with a lepton beam scattering off the transversely polarized proton target at COMPASS and compare with the corresponding experimental data. Due to the agreement between the theoretical calculation and the experimental data in SIDIS process, then we change the sign of the Sivers function applied in the SIDIS process to estimate the weighted $\frac{q_T}{M_p}\sin (\phi_S)$ asymmetry
in Drell-Yan process with $\pi^-$ beam scattering off the transversely polarized proton at COMPASS.

The rest of the paper is organized as follows.
In Sec.~\ref{sec:formalisms}, we provide the theoretical expressions of the weighted Sivers asymmetries in charged hadron produced SIDIS process and $\pi^- p$ Drell-Yan process.
In Sec.~\ref{sec:numerical}, we numerically estimate the weighted $\frac{P_{hT}}{zM_p}\sin (\phi_h-\phi_S)$ asymmetry in SIDIS process and the weighted $\frac{q_T}{M_p}\sin (\phi_S)$ asymmetry in $\pi^- p$ Drell-Yan process at COMPASS. We summarize the paper and discuss the results in Sec.~\ref{sec:conclusion}.

\section{FORMALISM OF THE WEIGHTED SIVERS ASYMMETRIES}
\label{sec:formalisms}
\subsection{ Weighted Sivers asymmetry in SIDIS process}
\begin{figure}
  \centering
  \includegraphics[width=0.6\columnwidth]{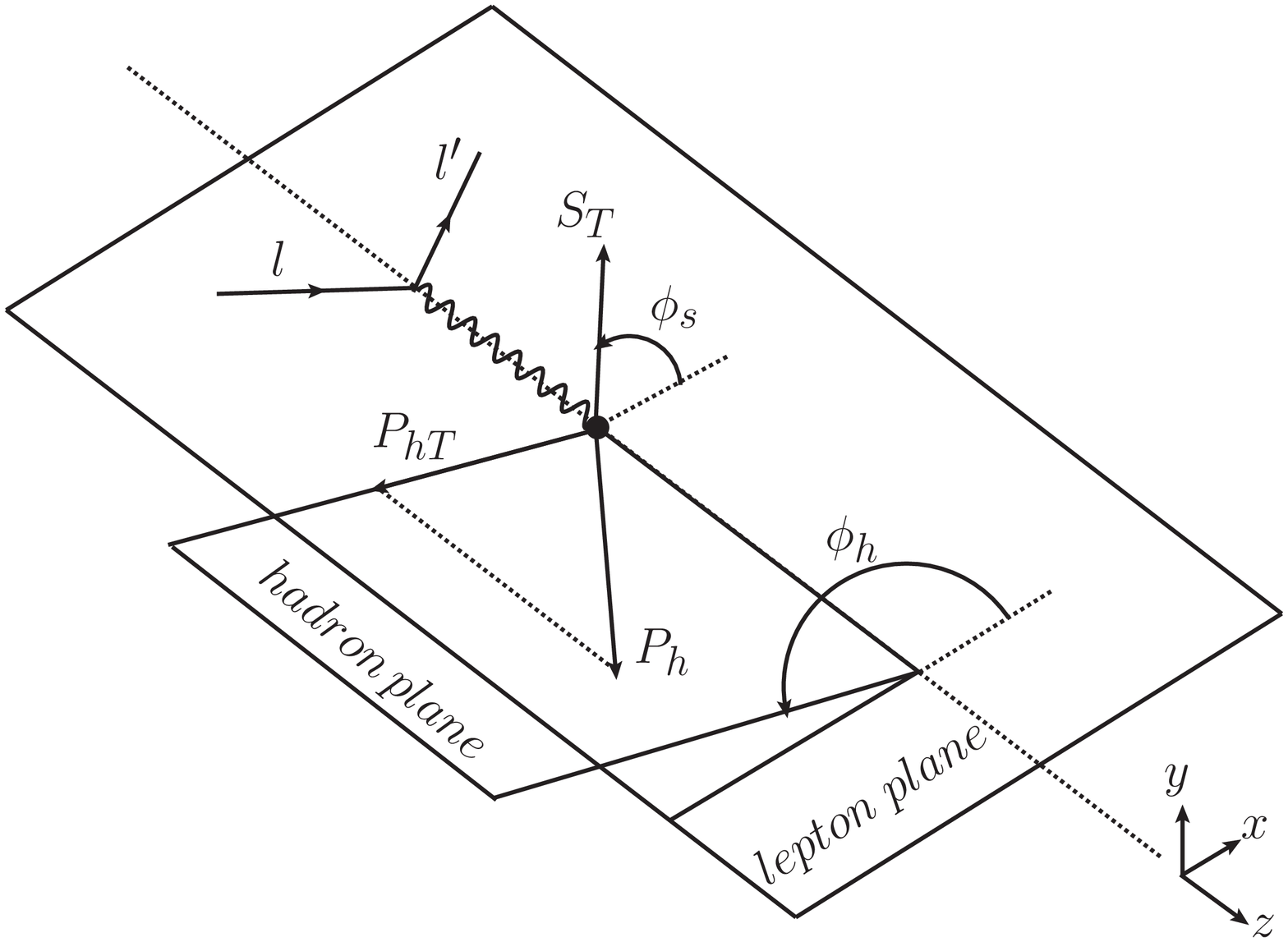}
  \caption{Reference frame of the transversely polarized SIDIS process.  }
  \label{fig:SIDIS}
\end{figure}

The process under study is the SIDIS process, particularly an unpolarized lepton beam scatters off the transversely polarized proton target to  produce an unpolarized charged hadron in the final-state:
\begin{equation}
\label{eq:sidis}
l(\ell)+p^\uparrow(P_p) \longrightarrow l(\ell^\prime)+h^{\pm} (P_h)+X(P_X),
\end{equation}
where $\ell$ and $\ell'$ denote the four-momenta of the incoming and outgoing leptons, respectively; $P_p$ and $P_h$ represent the four-momenta of the proton target $p$ and the final state hadron $h^{\pm}$, respectively; $\uparrow$ denotes that the proton target is transversely polarized.
$q=\ell-\ell^\prime$ denotes the four-momentum of the virtual photon with invariant mass squared $Q^2=-q^2$.
The reference frame of the process is depicted in Fig.~\ref{fig:SIDIS}, in which the momentum direction of the virtual photon defines the $z$ axis in accordance with the Trento conventions~\cite{Bacchetta:2004jz}.
$P_{hT}$ and $S_T$ are the component of $P_h$ and the proton spin vector $S$ that is transverse to the virtual photon momentum direction~($z$ axis).
$\phi_h$ denotes the azimuthal angle between the lepton and hadron planes which has the definition in the target rest frame as
\begin{align}
\label{eq:phih}
    \cos \phi_h=\frac{(\bm{\hat{q}}\times \bm{\ell})}{|\bm{\hat{q}}\times\bm{\ell}|}\cdot\frac{(\bm{\hat{q}}\times \bm{P_h})}{|\bm{\hat{q}}\times\bm{P_h}|}.
\end{align}
The azimuthal angle in any frames can be obtained by a boost along $\bm{\hat{q}}$, therefore, one can write the Lorentz invariant form of Eq.~(\ref{eq:phih}) into
\begin{align}
\label{eq:phih-LoInv}
    \cos \phi_h=-\frac{g^{\mu\nu}_\perp l_\mu P_{h\nu}}{|l_T||P_{hT}|},\quad
    \sin \phi_h=-\frac{\epsilon^{\mu\nu}_\perp l_\mu P_{h\nu}}{|l_T||P_{hT}|},
\end{align}
where $|l_T|=\sqrt{g^{\mu\nu}_\perp l_\mu l_\nu}$, $|P_{hT}|=\sqrt{g^{\mu\nu}_\perp P_{h\mu}P_{h\nu}}$.  $g^{\mu\nu}_\perp$ and ${\epsilon^{\mu\nu}_\perp}$ are the perpendicular projection tensors
\begin{align}
\label{eq:TranTensors}
    &g^{\mu \nu}_\perp=g^{\mu \nu}-\frac{q^\mu P_p^\nu+P_p^\mu q^\nu}{P_p\cdot q (1+\gamma^2)}+\frac{\gamma^2}{1+\gamma^2}(\frac{q^\mu q^\nu}{Q^2}-\frac{P_p^\mu P_p^\nu}{M_p^2}),\\
    &\epsilon^{\mu\nu}_\perp=\epsilon^{\mu\nu\rho\sigma}\frac{P_{p\rho} q_\sigma}{P_p\cdot q \sqrt{(1+\gamma^2)}},
\end{align}
with $\gamma = \frac {2M_px}{Q}$, $M_p$ being the proton mass and $x$ being the Bjorken variable $x=\frac {Q^2}{2P_p \cdot q}$.
$\phi_S$ stands for the azimuthal angle between the direction of $S_T$ and the lepton scattering plane which has the similar definition in Eqs.~(\ref{eq:phih}) and (\ref{eq:phih-LoInv}) with $P_h$ replaced by the covariant spin vector $S$ of the target proton.

Besides the Bjorken variable $x$ and $\gamma$, the following Lorentz invariants are introduced to express the differential cross section of the SIDIS process as well as the experimental observables
\begin{eqnarray}
y=\frac{P_p \cdot q}{P_p \cdot \ell},\quad z=\frac{P_p \cdot P_h}{P_p \cdot q}\,
,\quad s=(P_p+\ell)^2\,,
\end{eqnarray}
where $s$ is the center of mass energy squared of the $e$-$p$ system, $y$ is the inelasticity,
$z$ is the longitudinal momentum fraction of the final-state hadron.
The variables $x$ and $y$ are related to $Q^2$ through $xy=\frac{Q^2}{s}$~(actually, $xy=\frac{Q^2}{s-m_\ell^2-M_p^2}$, while the lepton mass and proton mass are neglected in general).

Assuming single-photon exchange, the SIDIS cross section can be expressed in terms of 18 structure functions~\cite{Bacchetta:2006tn} in a model-independent way.
Here, we only consider the terms relevant to the Sivers function, with other terms absorbed into the ellipsis~\cite{Bacchetta:2006tn,Bacchetta:2010si}
\begin{eqnarray}
\frac{d\sigma}{dx dy dz d\phi_S d\phi_h d\bm{P}^2_{hT}} =
\frac{\alpha^2}{xyQ^2}\,\frac{y^2}{2(1-\epsilon)}\left( 1+\frac{\gamma^2}{2x}\right)\,
\biggl[F_{UU}
+|\bm{S}_T|\,\sin (\phi_h-\phi_S)\,F_{UT}^{\sin (\phi_h-\phi_S)} + \ldots\biggr], \label{eq:cs}
\end{eqnarray}
where $\alpha=e^2/(4\pi)$ is the fine structure constant, and $\epsilon$ is the ratio of the longitudinal and transverse photon flux $\epsilon=\frac{1-y-\frac{1}{4}\gamma^2y^2}{1-y+\frac{1}{2}y^2+\frac{1}{4}\gamma^2y^2}$ such that the depolarization factor as $\frac{y^2}{2(1-\epsilon)}\approx(1-y+\frac{1}{2}y^2)$. $F_{UU}$ stands for the unpolarized structure function and $F_{UT}^{\sin\left(\phi_h -\phi_S\right)}$ represent the transverse spin-dependent structure function, with the subscripts $U$~(unpolarized) or $T$~(transverse polarized) denoting the polarization states of the beam (first subscript) and the target (second subscript).
The structure functions can be expressed as the following convolutions of the TMD PDFs and FFs~\cite{Bacchetta:2006tn}:
\begin{eqnarray}
F_{UU} &=& {\cal C} \biggl[ f_1 D_1 \biggr] \; ,  \\
F_{UT}^{\sin (\phi_h-\phi_S)} &=& {\cal C} \biggl[
-\frac{ \hat{\bm{h}}\cdot \bm{p}_{T}}
     {M_p} \, f_{1T}^{\perp}\,D_1\biggr] \;  \label{eq:conv_ut}
\end{eqnarray}
with the notation $\mathcal{C}$ representing the convolution of the transverse momentum:
\begin{equation}
\label{eq:note_C}
\mathcal {C}\bigl[ \omega f  D \bigr]
= x\sum_q e_q^2 \int d^2 \bm{p}_T  d^2 \bm{k}_T \delta^{(2)}\bigl(z\bm{p}_T + \bm{k}_T - \bm{P}_{hT} \bigr)\omega(\bm{p}_T,-\bm{k}_T/z)
f^q(x,\bm{p}_T^2)\,D^q(z,\bm{k}_T^2),
\end{equation}
where the sum runs over all the quark and antiquark flavors, $\hat{\bm{h}} = \bm{P}_{hT} / |\bm{P}_{hT}|$, $\bm{p}_{T}$ and $\bm{k}_{T}$ are the transverse momentum of quarks in the target proton and that of the final-state hadron relative to the fragmentation quark, respectively;
$\omega$ is a function of $\bm{p}_{T}$ and $\bm{k}_{T}$, 
$f_1(x,\bm{p}_T^2)$ and $f_{1T}^{\perp}(x,\bm{p}_T^2)$ are the unpolarized TMD PDF and the Sivers function, respectively; $D_1(z,\bm{k}_T^2)$ is the unpolarized FF.
 
Since the evolution effects of the TMD PDF and FF encoded by the Collins-Soper equation are complicated and can not be analytically solved without the parametrization of the nonperturbative evolution kernel~\cite{Collins:2011zzd},
for simplicity, we assume Gaussian form for TMD PDF and FF~\cite{Anselmino:2008jk, Anselmino:2007fs, Anselmino:2013vqa, Bradamante:2017yia, Lefky:2014eia}:
\begin{align}
f(x,\bm{p}_{T}^2) = f(x)\,\frac{1}{\pi  \langle p_T^2\rangle }\,\exp\left(-\frac{\bm{p}_{T}^2}{ \langle p_T^2\rangle }\right)\;,\nonumber\\
D(z,\bm{k}_{T}^2) = D(z)\,\frac{1}{\pi  \langle k_T^2\rangle }\,\exp\left(-\frac{\bm{k}_{T}^2}{ \langle k_T^2\rangle }\right)\;.
\label{eq:f1}
\end{align}
Here, $f(x)$ and $D(z)$ are the collinear PDF and FF that depend on $Q^2$, that is, they follow the Dokshitzer-Gribov-Lipatov-Altarelli-Parisi (DGLAP) evolution.
$\langle p_T^2\rangle$ and $\langle k_T^2\rangle$ are the average values of $\bm{p}_{T}^2$ and $\bm{k}_{T}^2$, which will be eliminated in the integral calculation over the intrinsic transverse momenta.
This Gaussian form of TMD PDF and FF is at tree level and is suitable to describe nonperturbative effects at small transverse momenta.

In order to simplify the complicated TMD evolution effects as well as the convolution in the transverse momentum space, we will consider the transverse momentum weighted asymmetry,
which can be expressed as the simple product of the collinear PDFs and FFs.
Following Refs.~\cite{Boer:1997nt,Bacchetta:2010si},  weighted asymmetry with proper weighting function $W$ is defined as:
\begin{eqnarray}
A_{UT}^W =\frac{\int d^2 \bm{P}_{hT} W  F_{UT}}{\int d^2 \bm{P}_{hT}  F_{UU}}.
\end{eqnarray}
After choosing the weighting function $W=\frac{P_{hT}}{zM_p}$, the weighted Sivers asymmetry can be written as
\begin{eqnarray}
A_{UT}^{\sin (\phi_h-\phi_S)\,\frac{P_{hT}}{zM_p}}  =\frac{\int d^2 \bm{P}_{hT} \frac{P_{hT}}{zM_p}  F_{UT}^{\sin\left(\phi_h -\phi_S\right)}}{\int d^2 \bm{P}_{hT}  F_{UU}}.
\end{eqnarray}
with the denominator $F_{UU}$ being
\begin{align}
 \label{eq:sidis-fenmu-w}
		\int d^2 \bm{P}_{hT}  F_{UU} 
		=& \int d^2  \bm{P}_{hT} \, {\cal C}  \bigg[f_1 \; D_1 \bigg]  \nonumber \\
		=& \ x\sum_q e_q^2 \int d^2 \bm{P}_{hT} d^2 \bm{p}_T  d^2 \bm{k}_T \delta^{(2)}\bigl(z\bm{p}_T + \bm{k}_T - \bm{P}_{hT} \bigr)
        f_1(x,\bm{p}_T^2) D_1(z,\bm{k}_T^2) \nonumber\\
        =& \sum_q e_q^2	\left [xf_1(x)\,D_1(z)\right].
\end{align}
Similarly, the $P_{hT}$-weighted spin-dependent structure function $\frac{P_{hT}}{zM_p}F_{UT}^{\sin\left(\phi_h -\phi_S\right)}$ can be written as
\begin{align}
 \label{eq:sidis-fenzi-w}
		\int d^2 \bm{P}_{hT}  \frac{P_{hT}}{zM_p}F_{UT}^{\sin\left(\phi_h -\phi_S\right)}
		=& \int d^2  \bm{P}_{hT} \frac{P_{hT}}{zM_p}
		{\cal C} \biggl[-\frac{\hat{\bm{h}}\cdot \bm{p}_{T}} {M_p} \, f_{1T}^{\perp}\,D_1\biggr]  \nonumber \\
		=&~x\sum_q e_q^2 \int d^2 \bm{P}_{hT}\frac{P_{hT}}{zM_p} d^2 \bm{p}_T  d^2 \bm{k}_T \delta^{(2)}\bigl(z\bm{p}_T + \bm{k}_T - \bm{P}_{hT} \bigr) \frac{-\hat{\bm{h}}\cdot \bm{p}_{T}} {M_p}
        f_{1T}^{\perp}(x,\bm{p}_T^2) D_1(z,\bm{k}_T^2) \nonumber\\
        =&~2\sum_q e_q^2
		\bigl[xf_{1T}^{\perp(1)}(x)\,D_1(z)\bigr],
\end{align}
where $f_{1T}^{\perp(1)}(x)$ is the first $\bm{p}_T$-moment of the Sivers function:
\begin{eqnarray}
f_{1T}^{\perp(1)}(x)=\int {d}^{2}\bm{p}_T\frac{\bm{p}_T^2}{2M_p^2} f_{1T}^{\perp}(x,\bm{p}_T^2).
\end{eqnarray}
In SIDIS process, $P_{hT}$-weighted Sivers asymmetry with weighting function $W=\frac{P_{hT}}{zM_p}$ can be rewritten as
\begin{eqnarray}
A_{UT}^{\sin (\phi_h-\phi_S)\,\frac{P_{hT}}{zM_p}}  = 2\,
\frac{\sum_q e_q^2	\bigl[xf_{1T}^{\perp(1)}(x)\,D_1(z)\bigr]}
    {\sum_q e_q^2	\bigl[xf_1(x)\,D_1(z)\bigr]} \; .
\label{eq:asymmetry in SIDIS}
\end{eqnarray}

\subsection{ Weighted Sivers asymmetry in Drell-Yan process}
\begin{figure}
  \centering
  \includegraphics[width=0.6\columnwidth]{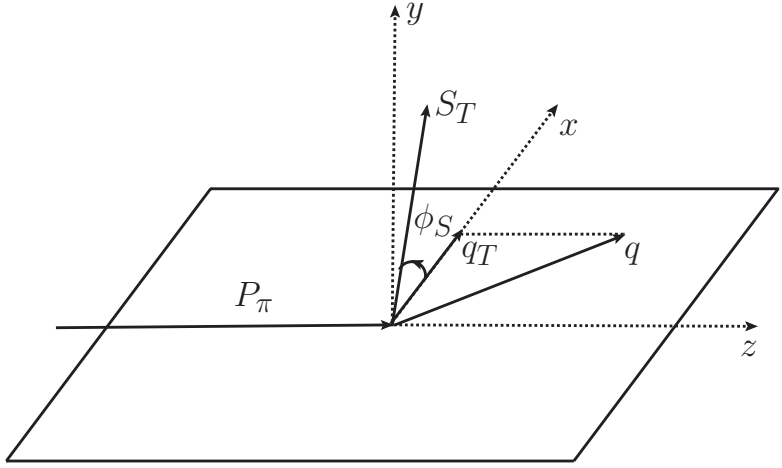}
  \caption{Target rest frame of the  $\pi^- p$ Drell-Yan process.}
  \label{DY}
\end{figure}

In this subsection, we will set up the theoretical framework of the transverse momentum weighted Sivers asymmetry in Drell-Yan process in which a $\pi^-$ beam scatters off a transversely polarized proton target:
\begin{align}
\pi^-(P_{\pi})+p^\uparrow(P_p)\longrightarrow \gamma^*(q)+X(P_X) \longrightarrow l^+(\ell)+l^-(\ell')+X(P_X),
\end{align}
where $P_{\pi}$, $P_p$, and $q$ represent the four-momenta of the pion beam $\pi^-$, the proton target $p$, and the virtual photon $\gamma^*$, respectively.
One should note that $q$ is a time-like vector different from that in the SIDIS process, namely, $Q^2=q^2>0$. $Q^2$ is the invariant mass square of the final-state dilepton.
The target rest frame of the process under study is shown in Fig.~\ref{DY}, in which the target proton is at rest and the momentum direction of the incident pion defines $z$ axis.
$q_T$ stands for the transverse component of $q$.
$S_T$ is the transverse spin of the target proton, and $\phi_S$ denotes the angle between the direction of $S_T$ and the hadron plane.

The following Lorentz invariants are introduced to express the differential cross section of the Drell-Yan process as well as the experimental observables:
\begin{align}
&x_\pi=\frac{Q^2}{2P_\pi\cdot q},\quad x_p=\frac{Q^2}{2P_p\cdot q},\quad x_F=2q_L/s=x_\pi-x_p,\nonumber\\
&s=(P_{\pi}+P_p)^2,\quad \tau=Q^2/s=x_\pi x_p,\quad y=\frac{1}{2}\mathrm{ln}\frac{q^+}{q^-}=\frac{1}{2}\mathrm{ln}\frac{x_\pi}{x_p},
\end{align}
where $s$ represents the center of mass energy squared of the $\pi^- p$ system.
$x_\pi$ and $x_p$ are the Bjorken variables of the pion and proton, respectively; $x_F$ is the Feynman $x$ variable with $q_L$ being the longitudinal momentum of the virtual photon; $y$ is the rapidity of the dilepton.

At leading order, the cross section of the transversely polarized $\pi^- p$ Drell-Yan process can be written as \cite{Arnold:2008kf, COMPASS:2010shj}:
\begin{align}
\frac{d \sigma}{ d q^4 d \Omega}=\frac{\alpha^2_{em}}{Fq^2}\,
&\biggl[  (1+\cos^2 \theta) F_{UU}
+|\bm{S}_T|  (1+\cos^2 \theta) \sin \phi_S \, F_{UT}^{\sin \phi_S} +\cdots   \biggr],
\end{align}
where the solid angle $\Omega$ specifies the orientation of the dilepton.
$F$ represents the flux of incoming hadrons.
$F_{UU}$ and $F_{UT}^{\sin \phi_S}$ are the relevant structure functions.
Using the notation  $\cal C $,
\begin{align}
  \mathcal{C}\bigl[\omega \, f_{q/\pi} \, f_{\bar{q}/p}\bigr] = \frac{1}{N_c} \; \sum_q e_q^2\int d^2\bm{p}_{T\pi} \, d^2\bm{p}_{Tp} \, \delta^{(2)}(\bm{q}_T- \bm{p}_{T\pi}-\bm{p}_{Tp}) \,
    \omega \, f_{q/\pi}(x_\pi,\bm{p}_{T\pi}^2)f_{\bar{q}/p}(x_p,\bm{p}_{Tp}^2)\,,
\end{align}
the two structure functions can be expressed as the following convolutions~\cite{Arnold:2008kf}:
\begin{align}
 &F_{UU} = {\cal C}  \bigg[f_{1,q/\pi} \; f_{1,\bar{q}/p} \bigg], 	\\
 &F_{UT}^{\sin \phi_S} ={\cal C} \bigg[ \frac{\hat{\bm{h}}\cdot \bm{p}_{Tp}}{M_p} \;
     f_{1,q/\pi} \; f_{1T,\bar{q}/p}^{\perp } \; \bigg],
\end{align}
where $\hat{\bm{h}} = \bm{q}_{T} / |\bm{q}_{T}|$.
$M_p$ is the proton mass.
$f_{1,\bar{q}/p}(x_p,\bm{p}_{Tp}^2)$ and $f_{1T,\bar{q}/p}^{\perp}(x_p,\bm{p}_{Tp}^2)$ are the unpolarized TMD PDF and the Sivers function of proton, which depend on the Bjorken variable $x_p$ and the transverse momentum $\bm{p}_{Tp}$ of the quark inside proton.	
$f_{1,q/\pi}(x_\pi,\bm{p}_{T\pi}^2)$ is the unpolarized TMD PDF of pion that depend on $x_\pi$ and $\bm{p}_{T\pi}$.

After choosing the weighting function $W=\frac{q_{T}}{M_p}$, the weighted Sivers asymmetry in Drell-Yan process can be written as:
\begin{eqnarray}
A_{UT}^{ \sin (\phi_S)\frac{q_T}{M_p}} =\frac{\int d^2 \bm{q}_T \frac{q_T}{M_p}  F_{UT}^{\sin\phi_S}}{\int d^2 \bm{q}_T  F_{UU}}.
\end{eqnarray}
with the denominator being
\begin{align}
 \label{eq:dy-fenmu-w}
		\int d^2 \bm{q}_T  F_{UU}=& \int d^2  \bm{q}_T \, {\cal C}  \bigg[f_{1,q/\pi} \; f_{1,\bar{q}/p} \bigg]  \nonumber \\
		=& \frac{1}{N_c}  \sum_q e_q^2\int d^2 \bm{q}_T \,  d^2\bm{p}_{T\pi} \, d^2\bm{p}_{Tp} \, \delta^{(2)}(\bm{q}_T- \bm{p}_{T\pi}-\bm{p}_{Tp}) \,
                f_{1,q/\pi}(x_\pi,\bm{p}_{T\pi}^2)f_{1,\bar{q}/p}(x_p,\bm{p}_{Tp}^2) \nonumber \\
        =&  \frac{1}{N_c} \sum_q e_q^2
					\bigl[	f_{1,q/\pi}(x_\pi) \,
							f_{1,\bar{q}/p}(x_p)\bigr].
\end{align}
For the spin dependent structure function $F_{UT}^{\sin\left(\phi_S\right)}$, we have
\begin{align}
 \label{eq:dy-fenzi-w}
		\int d^2 \bm{q}_T \frac{q_T}{M_p}  F_{UT}^{\sin\phi_S}
		=& \int d^2  \bm{q}_T \frac{q_T}{M_p}  \, {\cal C}  \bigg[ \frac{\hat{\bm{h}}\cdot \bm{p}_{Tp}}{M_p} \;
     f_{1,q/\pi} \; f_{1T,\bar{q}/p}^{\perp } \; \bigg]  \nonumber \\
		=& \frac{1}{N_c}  \sum_q e_q^2\int d^2 \bm{q}_T \,\frac{q_T}{M_p} \, d^2\bm{p}_{T\pi} \, d^2\bm{p}_{Tp} \, \delta^{(2)}(\bm{q}_T- \bm{p}_{T\pi}-\bm{p}_{Tp}) \,
               \frac{\hat{\bm{h}}\cdot \bm{p}_{Tp}}{M_p} \; f_{1,q/\pi}(x_\pi,\bm{p}_{T\pi}^2)f^\perp_{1T,\bar{q}/p}(x_p,\bm{p}_{Tp}^2) \nonumber \\
        =&  2\,\frac{1}{N_c} \sum_q e_q^2
					\bigl[	f_{1,q/\pi}(x_\pi) \,
							f_{1T,\bar{q}/p}^{\perp(1)}  (x_p)\bigr].
\end{align}

In Drell-Yan process, the $q_{T}$-weighted Sivers asymmetry with weighting function $W=\frac{q_{T}}{M_p}$ can be rewritten as:
\begin{align}
 \label{eq:asymmetry2}
		A_{UT}^{ \sin (\phi_S)\frac{q_T}{M_p}}
		&=  2\frac{\sum_q e_q^2
                    \bigl[  f_{1,q/\pi}(x_\pi) \,
	       f_{1T,\bar{q}/p}^{\perp(1)}  (x_p) \bigr]}        {\sum_q e_q^2
                    \bigl[  f_{1,q/\pi}(x_\pi) \,
	       f_{1,\bar{q}/p}  (x_p) \bigr]} .	
\end{align}

\section{NUMERICAL CALCULATION}
\label{sec:numerical}

Based on the above formalism, we present the numerical calculation for the weighted Sivers asymmetries in charged hadron produced SIDIS process and the $\pi^- p$ Drell-Yan process at the kinematical regions of COMPASS, and compare the numerical results with the SIDIS data~\cite{COMPASS:2018ofp} and Drell-Yan preliminary data~\cite{Longo:2019bih}.

To do this, the collinear unpolarized PDF of the proton $f_{1,p}(x_p)$ and unpolarized FF of the final-state hadron $D_1^q(z)$ are adopted from the CT10 parametrization~\cite{Lai:2010vv} and the DSS parametrization~\cite{deFlorian:2007aj}, respectively. For the unpolarized PDF of the pion $f_{1,\pi}(x_\pi)$, we use the SMRS parametrization~\cite{Sutton:1991ay}.

As mentioned above, one needs the information of the first $\bm{p}_T$-moment of proton Sivers function $f_{1T}^{\perp(1)}(x)$ in both SIDIS and Drell-Yan processes. $f_{1T}^{\perp(1)}(x)$ and the usual Qiu-Sterman function $T_{q,F}(x,x)$ have the following relation~\cite{Sun:2013hua,Kang:2011mr}:
\begin{align}
T_{q,F}(x,x)=\int {d}^{2}\bm{p}_T\frac{p_T^2}{M_p} f_{1T}^{\perp}(x,\bm{p}_T^2)=2M_p f_{1T}^{\perp(1)}(x). \label{eq:qs_moment}
\end{align}
In Ref.~\cite{Echevarria:2014xaa}, the authors have extracted Qiu-Sterman function by using the preliminary SIDIS data from HERMES~\cite{HERMES:2009lmz}, COMPASS~\cite{COMPASS:2008isr, COMPASS:2012dmt}, and JLab~\cite{JeffersonLabHallA:2011ayy} on Sivers asymmetry.
Assuming the Qiu-Sterman function is proportional to the unpolarized PDF $f_{1,p}(x_p)$, we adopt the parametrization~\cite{Echevarria:2014xaa} for $T_{q,F}(x, x)$ at the initial scale $Q^2_0=2.4~\rm{GeV}^2$:
\begin{align}
T_{q, F}(x, x) = N_q \frac{(\alpha_q+\beta_q)^{(\alpha_q+\beta_q)}}{\alpha_q^{\alpha_q} \beta_q^{\beta^q}}
x^{\alpha_q} (1-x)^{\beta_q} f_{1,p}(x_p),
\label{eq:T_qF}
\end{align}
and the values of the free parameters in global fit are~\cite{Echevarria:2014xaa}:
\begin{align}
 &\alpha_u =1.051_{-0.180}^{+0.192},\quad
 \alpha_d = 1.552_{-0.275}^{+0.303}, \quad
 \alpha_{sea} = 0.851_{-0.305}^{+0.307},\quad
  \beta = 4.857_{-1.395}^{+1.534},  \nonumber \\
 &N_{u}=0.106_{-0.009}^{+0.011}, \quad
 N_{d}=-0.163_{-0.046}^{+0.039}, \quad
 N_{s}=0.103_{-0.604}^{+0.548}, \nonumber \\
& N_{\bar{u}}=-0.012_{-0.020}^{+0.018}, \quad
 N_{\bar{d}}=-0.105_{-0.060}^{+0.043}, \quad
 N_{\bar{s}}=-1.000\pm1.757.
\label{eq:fits}
\end{align}
The extraction can give a rather good description for all the corresponding data of the differential cross sections.
We assume that Qiu-Sterman function follows the proportional relation with $f_{1,p}(x)$ in all the energy region.
Note that the Sivers function applied in Drell-Yan process has a opposite sign with respect to the one extracted from SIDIS data.

The kinematical ranges in SIDIS process covered by COMPASS are as follows~\cite{COMPASS:2018ofp, Bradamante:2017yia}:
\begin{align}
&0.004<x<0.7,\quad 0.1<y<0.9, \quad z>0.2, \nonumber    \\
&Q^{2}>1\ \mathrm{GeV}^2 \;, \quad   5\ \mathrm{GeV} < W < 18\ \mathrm{GeV},\quad s=301~\mathrm{GeV}^2.
\end{align}
where $W^2=(P_p+q)^2\approx \frac{1-x}{x}Q^2$  is the invariant mass squared of the virtual photon-proton system. As for the $\pi^- p$ Drell-Yan process at COMPASS,
we adopt the following kinematical cuts:
\begin{align}
&0.05<x_p<0.4,\quad 0.05<x_\pi<0.9, \quad
 -0.3<x_F<1,\nonumber    \\
&4.3\ \mathrm{GeV}<Q<8.5\ \mathrm{GeV},\quad s=357~\mathrm{GeV}^2.
\label{eq:cuts}
\end{align}

\begin{figure}
  \centering

  \includegraphics[width=0.4\columnwidth]{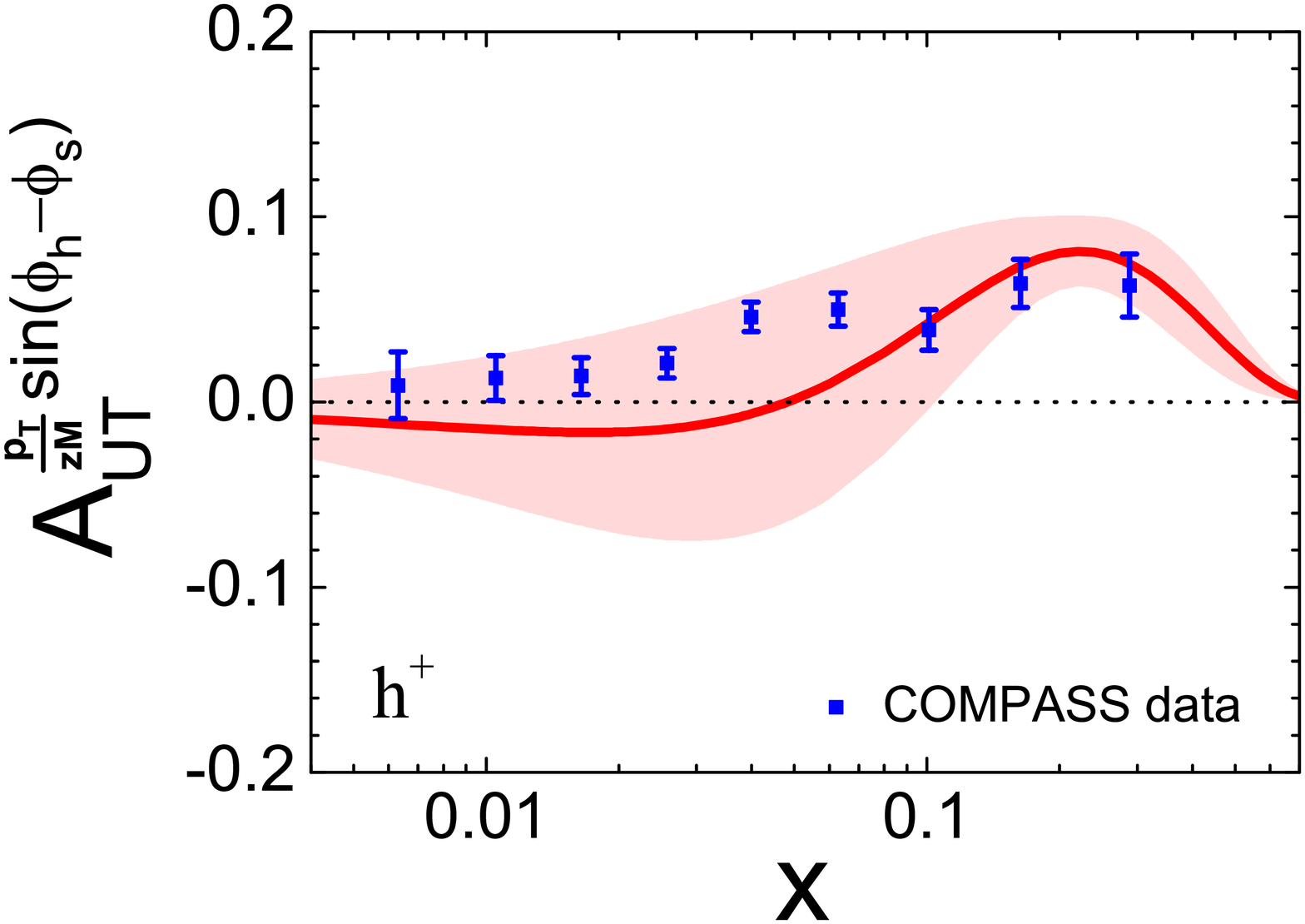}
  \includegraphics[width=0.4\columnwidth]{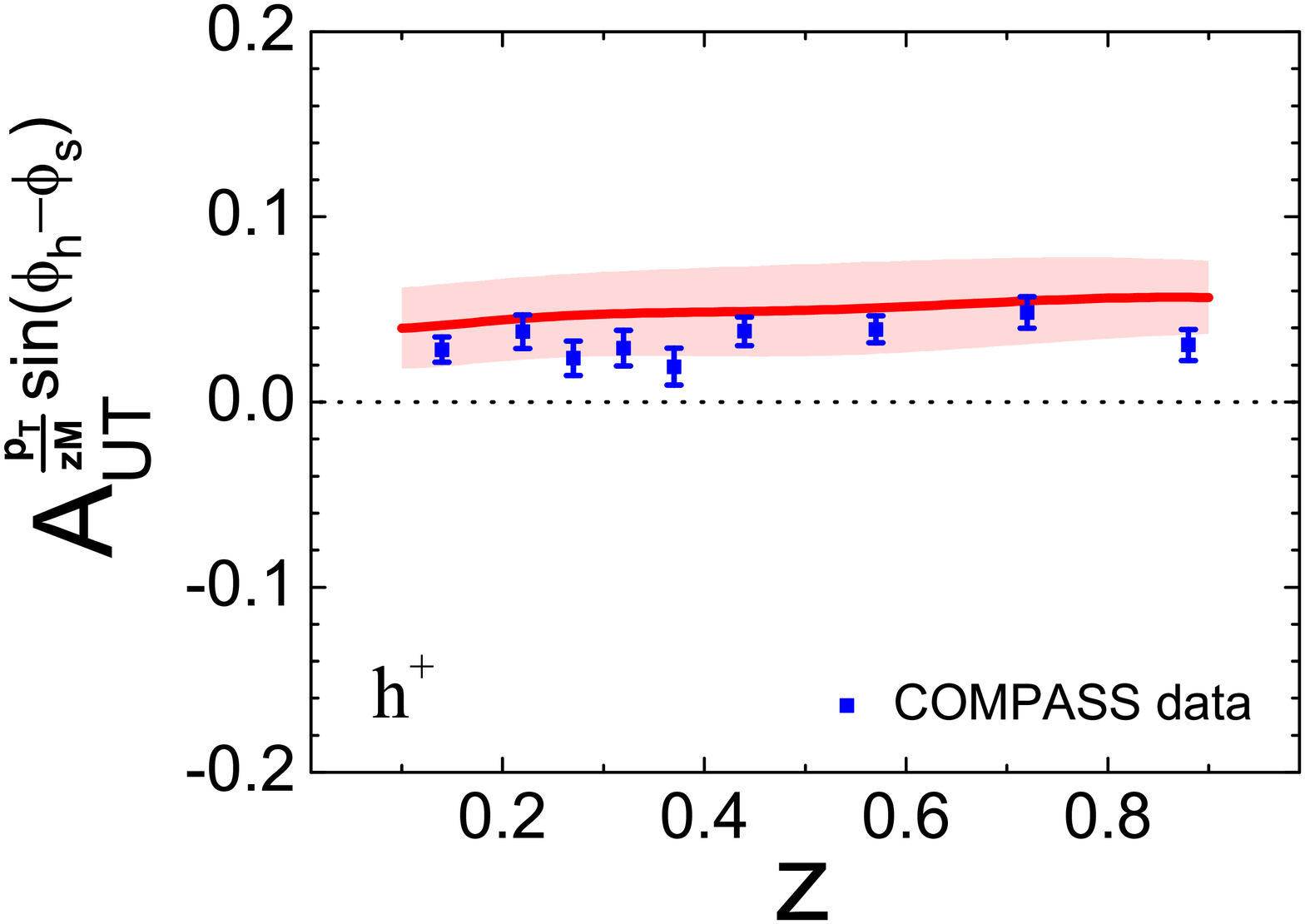}
  \includegraphics[width=0.4\columnwidth]{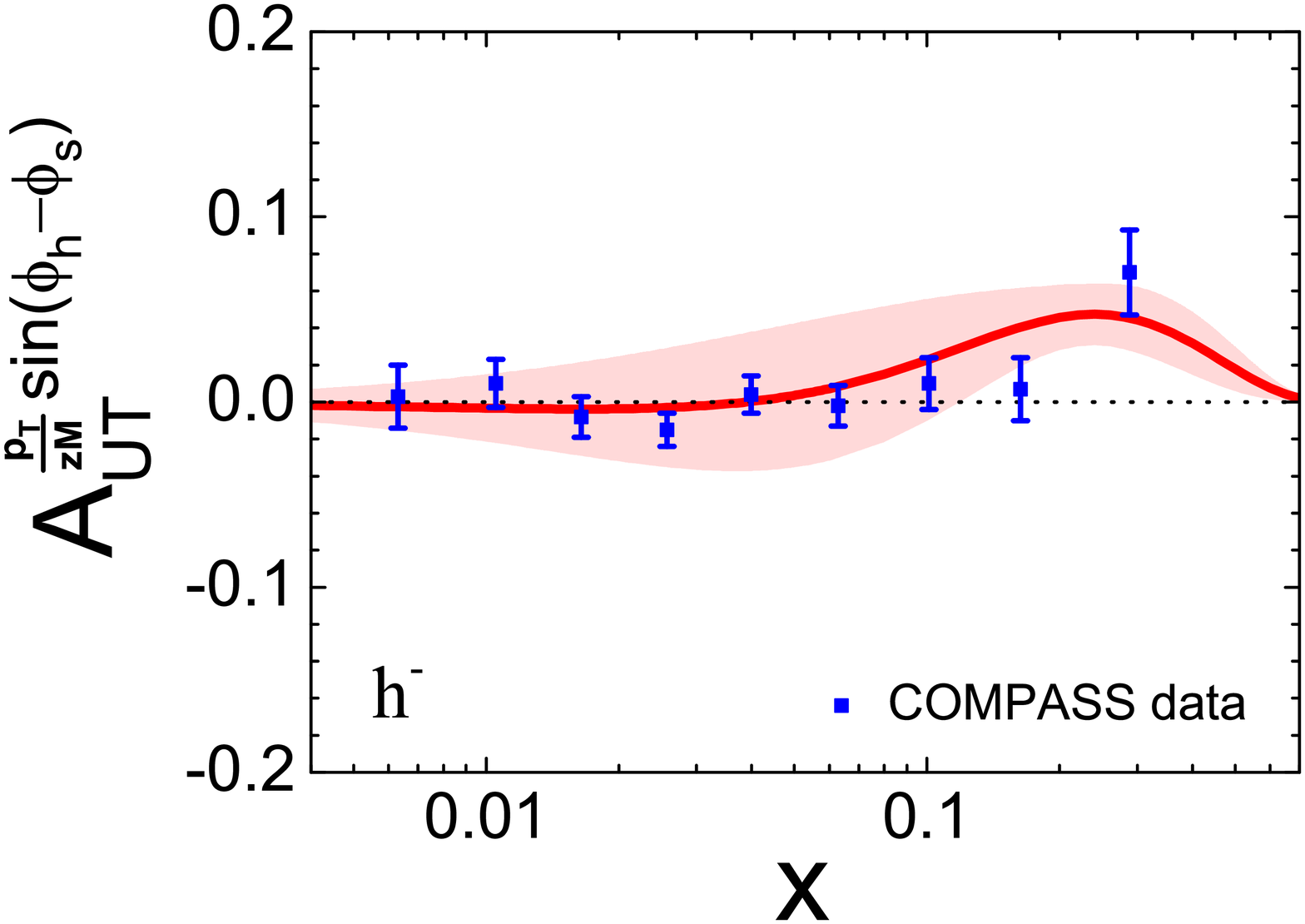}
  \includegraphics[width=0.4\columnwidth]{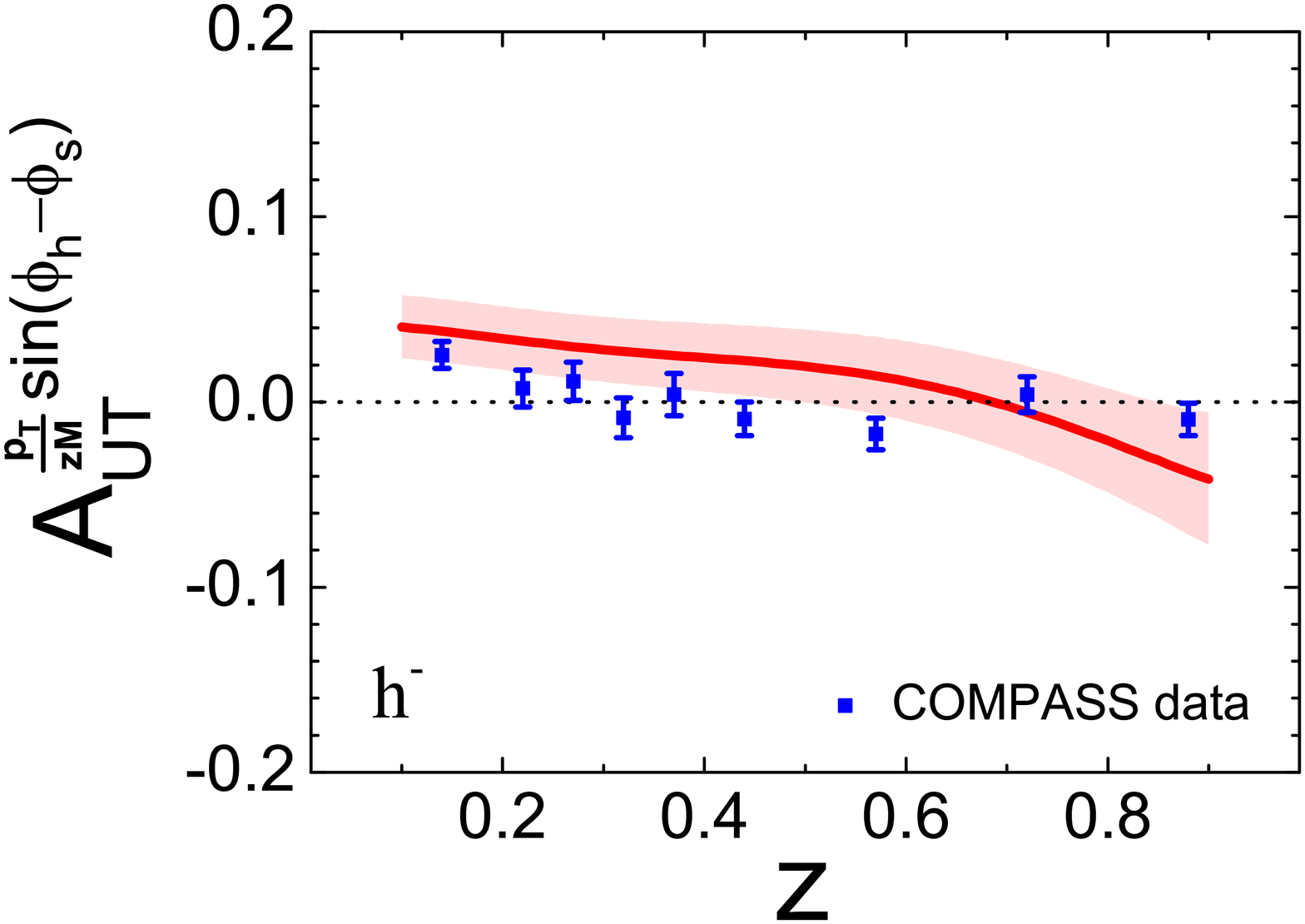}
   \caption{The weighted Sivers asymmetry of charged hadron produced in SIDIS process. The solid squares with error bars represent the COMPASS data for comparison~\cite{COMPASS:2018ofp}.}
  \label{fig:asy_SIDIS}
\end{figure}

We apply Eq.~(\ref{eq:asymmetry in SIDIS}) to calculate the $\frac{P_{hT}}{zM_p}$-weighted Sivers asymmetry of charged hadron produced in SIDIS process and plot the results in Fig.~\ref{fig:asy_SIDIS}. The upper two panels show the asymmetry for $h^+$ production; the lower panels give the results for $h^-$ production.
The left and right panels show the asymmetry as functions of $x$ and $z$, respectively. In each case the other kinematical variables are integrated out.
In Fig.~\ref{fig:asy_SIDIS}, the solid lines correspond to the results from the central values of the parameters in the parametrization of $T_{q,F}(x, x)$,
and the shaded areas represent the uncertainty bands due to the uncertainties of the parameters in Eq.~(\ref{eq:fits}).
As a comparison, we also show the
experimental data measured by the COMPASS Collaboration~\cite{COMPASS:2018ofp}.
As shown in Fig.~\ref{fig:asy_SIDIS}, our results are in agreement with the COMPASS data within errors.

\begin{figure}
  \centering
  \includegraphics[width=0.4\columnwidth]{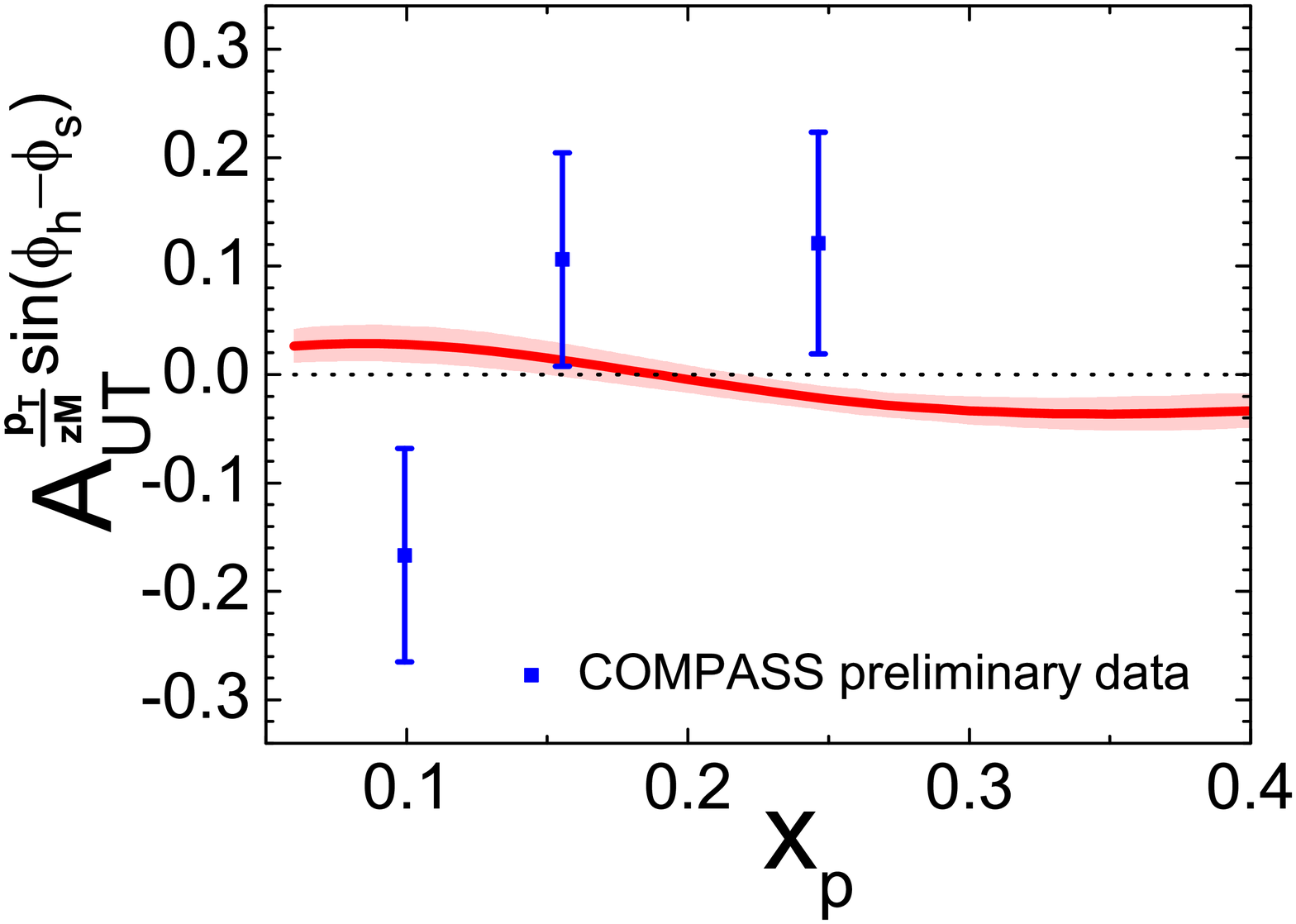}
  \includegraphics[width=0.4\columnwidth]{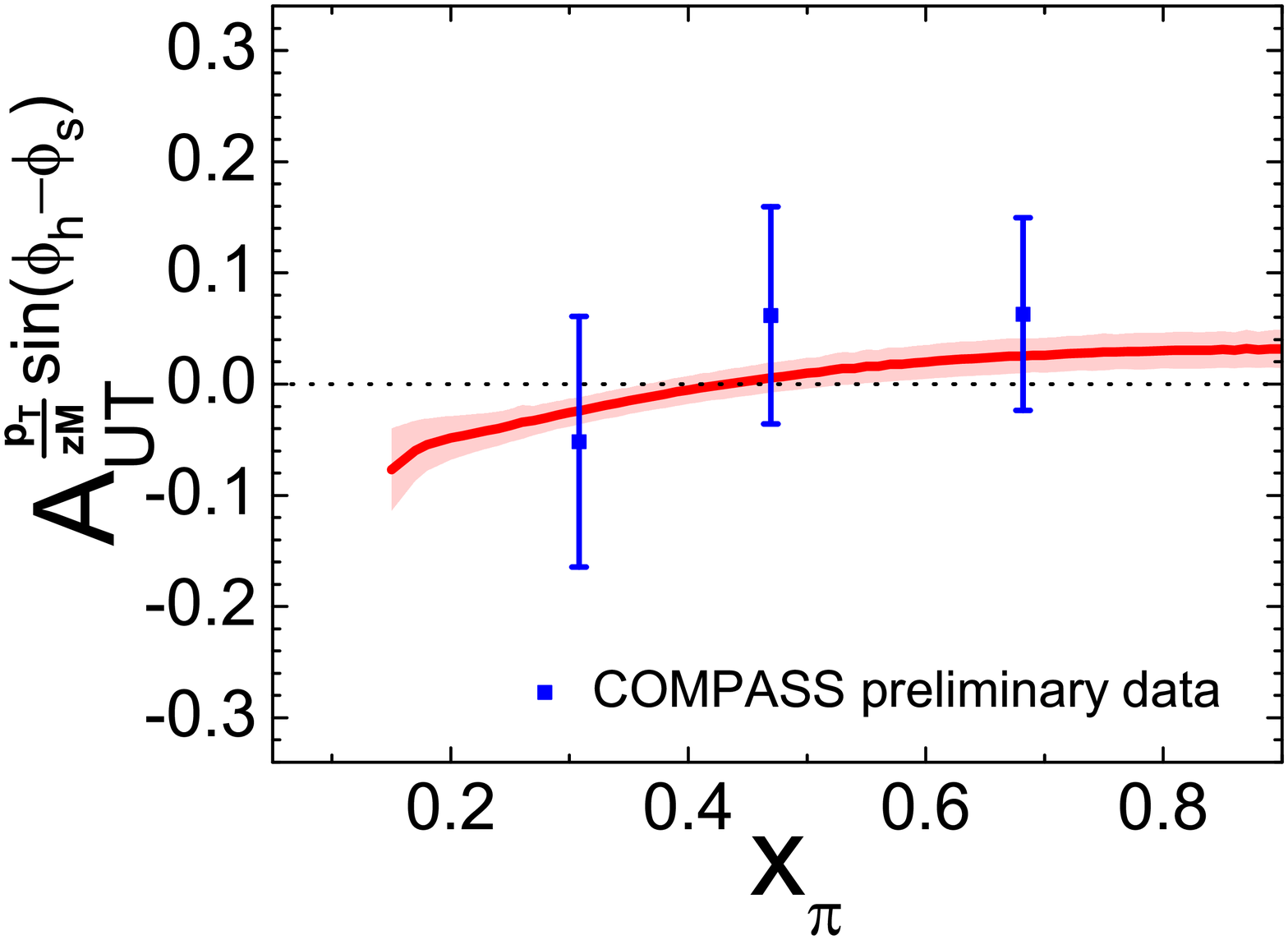}
  \includegraphics[width=0.4\columnwidth]{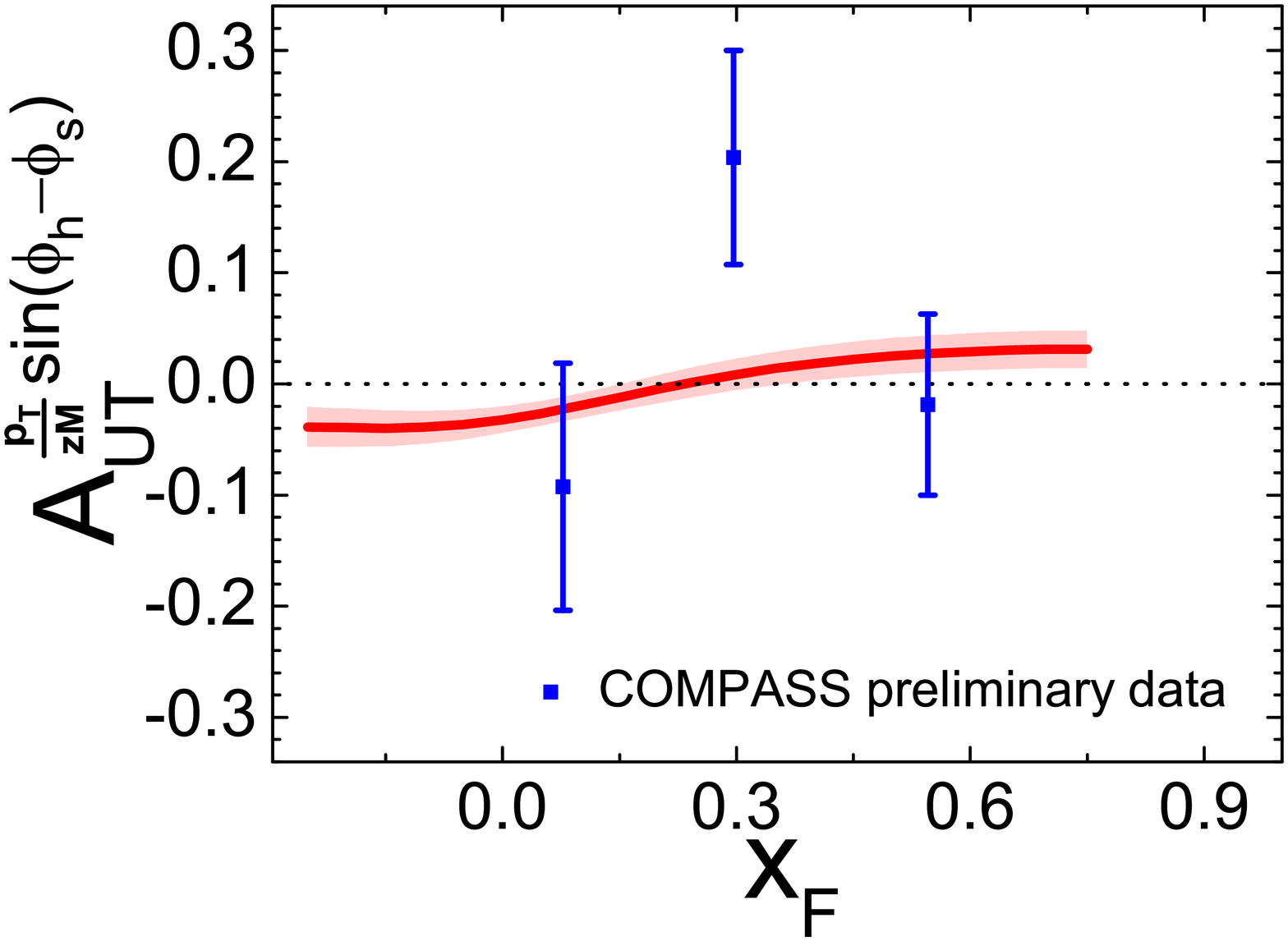}
  \includegraphics[width=0.4\columnwidth]{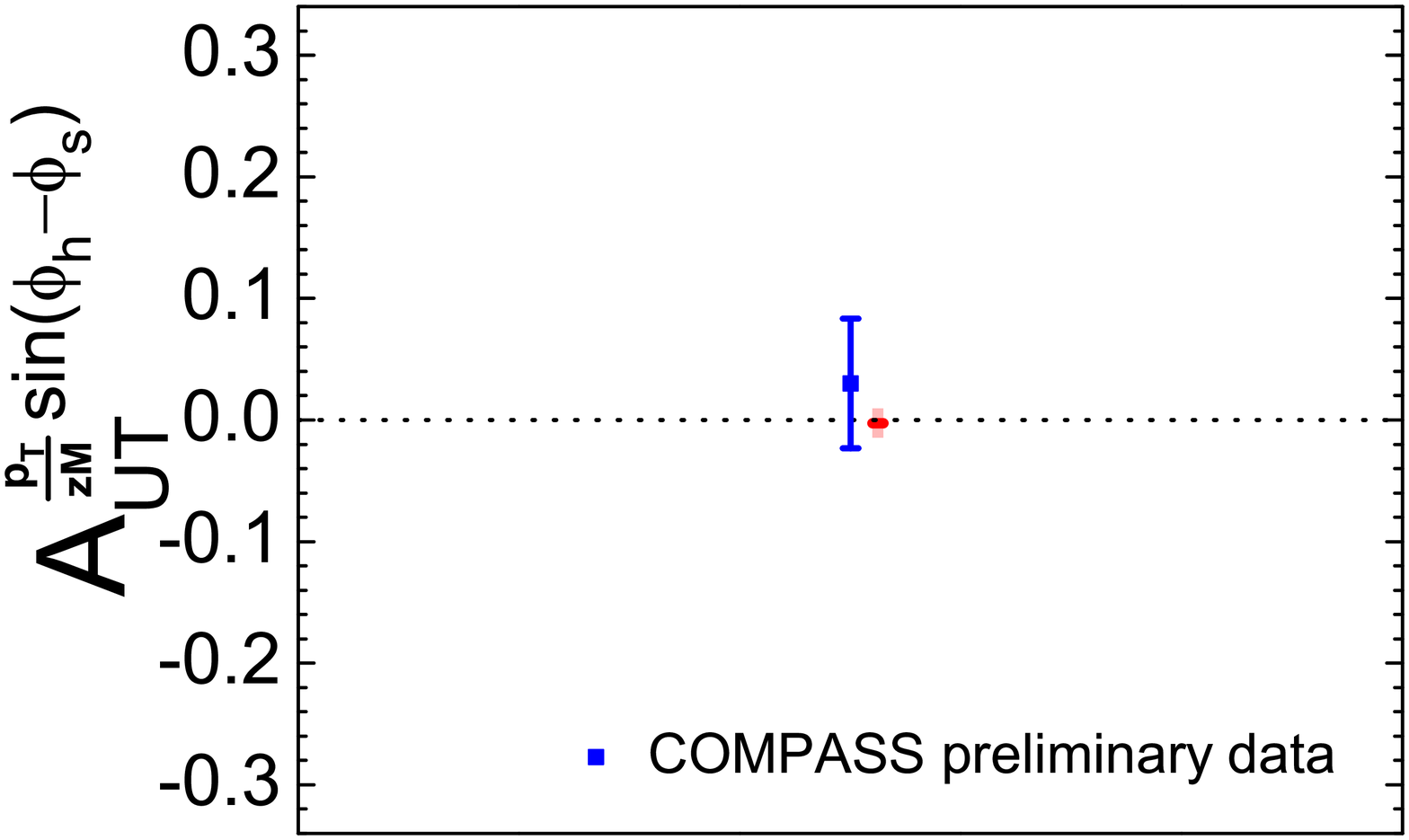}
   \caption{The weighted Sivers asymmetry in $\pi^- p$ Drell-Yan process. The solid squares with error bars are COMPASS preliminary data~\cite{Longo:2019bih}.}
  \label{fig:asy_DY}
\end{figure}

The agreement between our results and the COMPASS SIDIS data encourages us to further estimate the weighted Sivers asymmetry  $A_{UT}^{ \frac{q_T}{M_p}}\sin (\phi_S)$ in $\pi^- p$ Drell-Yan process by utilizing the relation $f_{1T}^{\perp \mathrm{(DY)}}=-f_{1T}^{\perp \mathrm{(SIDIS)}}$, which can be used to verify the sign-change of Sivers function. 
The corresponding results are plotted in Fig.~\ref{fig:asy_DY},
in which the four panels show the asymmetry as functions of $x_p$ (upper left), $x_\pi$ (upper right), $x_F$ (lower left) and the asymmetry integrated over the entire kinematical range (lower right).
In Fig.~\ref{fig:asy_DY}, the solid lines denote the central results, and the uncertainty bands are determined by the uncertainties of the parameters in Eq.~(\ref{eq:fits}).
The corresponding  preliminary COMPASS data with error bars~\cite{Longo:2019bih} are also shown in Fig.~\ref{fig:asy_DY}.
One can find the theoretical results of $x_\pi$-dependent asymmetry agree with the preliminary data.
Although there is difference between the preliminary data and the theoretical calculation of the tendency for the $x_p$-dependent asymmetry as well as the $x_F$-dependent asymmetry, the asymmetry integrated over the entire kinematical range still agrees with the preliminary COMPASS data within errors.
Compared to the shaded areas in Fig.~\ref{fig:asy_SIDIS}, the uncertainty bands of theoretical calculation for asymmetry in $\pi^- p$ Drell-Yan process are relatively narrow.
The reason may be that in Eq.~(\ref{eq:asymmetry2}) the contribution of sea quarks is small, thus the asymmetry is dominated by $2f_{1T,p}^{\perp(1)u(\mathrm{DY})}  (x_p)/f_{1,p}^u  (x_p)$. It indicates that only the uncertainty of the parameters for $u$ quark in Eq.~(\ref{eq:fits}) is the main source of the uncertainty bands, which means that the Sivers function of sea quark has not been well constrained. 

At present, we are still unable to conclude that the Sivers function has opposite sign in SIDIS process and in Drell-Yan process due to the limited Drell-Yan data with large errors. More and high precision experimental data are needed to verify the opposite sign of the Sivers function predicted by QCD and to constrain the sea quark Sivers function.

\section{CONCLUSION}
\label{sec:conclusion}
In this work, we have studied the weighted Sivers asymmetries in charged hadron produced SIDIS process and $\pi^- p$ Drell-Yan process.
The asymmetries are contributed by the first transverse-moment of the Sivers function, for which we have utilized the parameterization from literature.
The weighted $\frac{P_{hT}}{zM_p}\sin (\phi_h-\phi_S)$ asymmetry in SIDIS process and the weighted $\frac{q_T}{M_p}\sin (\phi_S)$ asymmetry in Drell-Yan process at the kinematics configurations of COMPASS SIDIS and Drell-Yan programs have been estimated, respectively. Our numerical estimate of the asymmetry in charged hadron produced SIDIS process is consistent with the COMPASS measurement.
However, due to the preliminary Drell-Yan data with large errors, our numerical results are still unable to demonstrate clearly the sign-change property of the Sivers function, and high precision experimental data are needed to clarify this point and to constrain the sea quark Sivers function.

\section{ACKNOWLEDGMENTS}
This work is partially supported by the NSFC (China) grants 11905187,11847217, 12150013. X. Wang is supported by the China Postdoctoral Science Foundation under Grant No.~2018M640680.

\end{document}